# Laser Micromachining of Coated $YBa_2Cu_3O_{6+x}$ Superconducting Thin Films

Principle Author: Kenneth E. Hix, Mound Laser & Photonics Center, Inc., P.O. Box 223, Miamisburg, OH 45343, 937-865-3041, 937-865-3680 fax, kenhix@mlpc.com

Matthew C. Rendina, Mound Laser & Photonics Center Inc. P. O. Box 223, Miamisburg, OH 45343, 937-865-3179, 937-865-3680 fax, matt@mlpc.com

James L. Blackshire, Materials & Manufacturing Directorate, Wright-Patterson Air Force Base, OH 45433, 937-255-0198, James.Blackshire@wpafb.af.mil

George A. Levin, Propulsion Directorate, Air Force Research Laboratory, Wright- Patterson Air Force Base, OH 45433, 937-255-4780, George.Levin@wpafb.af.mil

**ABSTRACT**

Over the last decade advances in high temperature superconducting (HTS) materials have generated a renewed optimism in the use of this technology in future high-power applications including motors, generators, and power transmission. The most promising superconducting material, $YBa_2Cu_3O_{6+x}$ (YBCO), has demonstrated current density capacity as high as $10^6$ $A/cm^2$ and is typically deposited epitaxially as a thin film on a buffered metallic substrate. The coated conductor architecture Ag/YBCO/Buffer/Ni-based substrate, as does other heteromaterials, introduces unique challenges to the laser micromachining process. In the present study, optimization of the laser micromachining process for machining grooves extending through the YBCO into the buffer layer is considered. A frequency tripled diode-pumped solid-state $Nd:YVO_4$ laser at 355 nm is used in conjunction with both fixed and scanning optical systems. The dependence of cut quality, dimension, and local damage to the YBCO layer on the laser processing parameters including focal spot size, pulse overlap, and fluence will be discussed. Preliminary experimental results indicate that laser micromachining may be used as a reliable process for patterning of superconducting tapes.

**1. INTRODUCTION**

High temperature superconducting (HTS) materials are typically characterized as ceramics or metals with a critical superconducting temperature above 25 Kelvin. Perhaps the most promising HTS ceramic material is yttrium barium copper oxide, $YBa_2Cu_3O_{6+7-x}$ (YBCO), which exhibits a critical temperature at 90 K – well above the 77 K boiling temperature of liquid nitrogen. With recent advances in coated conductor production processes YBCO based superconducting technology may soon be economically viable for various DC and AC applications including power generation and transmission for which current densities as high as $10^6$ $A/cm^2$ have been demonstrated.

YBCO is deposited epitaxially as a thin film on a metal substrate. The most common deposition techniques include ion beam assisted deposition (IBAD), rolling assisted biaxially textured substrates (RABiTS), as well as pulsed laser deposition (PLD). In the superconducting state the current transport flows along the copper oxide planes within the orthorhombic YBCO crystal structure. Thus, the superconducting properties heavily depend on the crystal structure and grain orientation of the thin film. The YBCO coated conductor manufacturing methodology typically begins with a long length 1-cm wide textured nickel-based alloy substrate with a thickness of approximately 100 microns on which a buffer layer is first deposited. Various buffer layer architectures are used and may be composed of either a single or multiple thin film combinations of ceramic or metallic materials. Buffer layer thickness largely depends on the chosen material combination. A 1-1.5 micron thick film of YBCO is deposited on top of the buffer. Finally, a 5-10 micron thick silver coating is deposited. A cross section of this architecture is shown in Figure 1.

In high-power AC applications, such as generators or motors, the superconductor wires are subjected to AC magnetic field, which results in very large hysteretic losses – Joule heat generated by the induced AC electric field interacting with very large current density. Carr and Oberly [1] suggested that the hysteretic losses can be reduced by dividing the wide superconducting tape into many filaments. The amount of heat generated in such striated tape decreases proportionally to the width of the individual filament. This idea has been confirmed experimentally by Cobb et al. [2] on small (3 mm x 12 mm) samples of YBCO thin film deposited on LaO substrate. The striations were laser micromachined and the losses were shown to decrease with the filament width as expected. The next necessary step is to produce striated samples using 1 cm wide and up to 10 cm long coated conductors produced by IBAD and RABiTS manufacturing methods in order to evaluate the losses under more realistic conditions. As a result, micro-manufacturing processes for introducing striations in the coated conductors are desired.

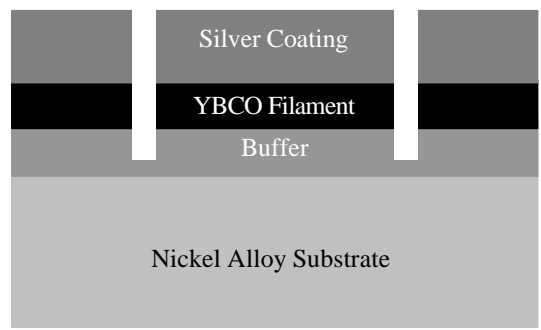

Figure 1. Typical coated conductor cross-section (desired cuts included).

Laser micromachining of the coated conductor architecture Ag/YBCO/Buffer/Ni-alloy substrate is considered in this study. Laser micromachining has several advantages in comparison to other material removal techniques such as chemical etching, focused ion beam (FIB), and mechanical scribing. Direct laser ablation of thin films is a cost effective and flexible processing technology, which lends itself well to rapid prototyping while requiring no masks or supplemental chemicals or processing steps. The laser micromachined patterns may also be quickly altered through CAD/CAM software control. This non-contact one step process is particularly advantageous to applications were the thin film patterning is a single step in a multi-step manufacturing scheme. The laser beam can be scanned rapidly and directed through a window into a chamber or to other inaccessible locations. Additionally, only the material intended for removal via laser ablation is exposed to the material removal agent. Thus laser micromachining is the perfect "bolt-on" material processing solution.

The coated conductor architecture, introduces unique challenges to the laser micromachining process. Variations in optical absorption properties as well as ablation thresholds may create a condition of preferential material removal. Additionally, the thermal conductivity, heat capacities, and phase change properties of each material within the layered structure may give rise to unique effects. The occurrence of a heat affected zone (HAZ) is of particular interest because its effect on the functionality of the coated conductor. The observed effects of machining this unique material set as well as the different laser micromachining methodologies employed are discussed.

## 2. EXPERIMENTAL

All laser micromachining experiments were performed at Mound Laser & Photonics Center Inc. (MLPC) employing a custom laser micromachining workstation capable of using multiple laser wavelengths and various optical arrangements. In this case a frequency tripled diode-pumped solid-state Nd:YVO$_4$ laser at 355 nm was utilized in all experiments. The laser pulse duration is 30-45 nanoseconds depending on pulse repetition rate, which may vary from 1 kHz to 100 kHz. The maximum average power is greater than 3.5 watts at 20 kHz. This type of laser demonstrates excellent pulse-to-pulse stability (<3.5% RMS at 20 kHz) and mode quality (TEM$_{00}$, M$^2$ <1.5).

Two different techniques were used to machine the desired cuts by scanning the laser beam across the surface of the coated conductor. The first approach was to fix the coated conductor in a stationary position using a vacuum stage and scan the laser beam across the surface using a galvanometer laser beam scanning head. The Scanlab HurrySCAN10-355 was used for all laser beam-scanning experiments. The focusing objective is mounted in the laser beam scanner housing and is easily removed and replaced. Two different focusing objectives having focal lengths of 103 mm and 53 mm were used to determine the effects of focal spot size on the laser micromachining process. The second approach was to focus the laser beam to a stationary point in space and translate the coated conductor in the focal plane of the laser beam. A simple plano-convex fused silica lens having a 100 mm focal length was used to focus the laser beam. The coated conductor was translated in the focal plane using a 4-axis Aerotech ATS1000 series translation stage.

In addition to optical microscopy and scanning electron microscopy, optical interferometry was performed on select samples for additional analysis. Optical interferometry techniques use the interference of light to measure surface topography features. Traditionally, the interference of two monochromatic beams of light is used, where one beam acts as a reference, and the other beam interacts with the material surface of interest. When recombined, variations in phase between these two light beams cause constructive and destructive interference to occur, which is manifested in light intensity changes as the local surface topography of the material surface changes. The local three-dimensional topographic features of an object can, therefore, be imaged through measurements of the interfering light intensity.

The optical interferometry technique used in these studies is termed 'white-light interferometry'. The technique involves the interference of reference and object beams of light, but instead of using a highly coherent, monochromatic light source, a partially coherent, polychromatic light source is used (see Figure 2). Because polychromatic light has a short coherence length, interference fringes appear only when the optical path length difference between the sample and reference surface is close to zero. As shown in Figure 1, the 'white-light' interference pattern involves an additional envelope function due to the polychromatic light that has a maximum fringe contrast at exactly zero path length difference between the sample and reference surfaces. By monitoring the fringe contrast as the sample is vertically scanned through focus, a precise measure of the local focal position can be measured, and the topographic features can be mapped.

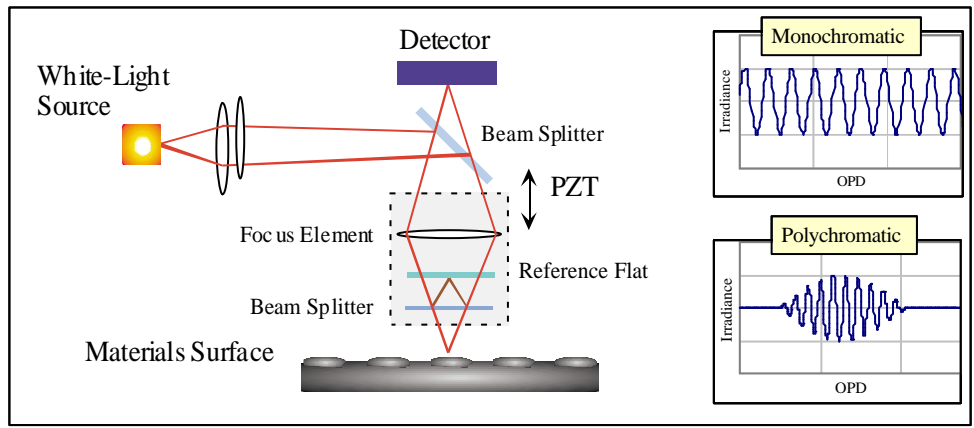

Figure 2. Schematic diagram of white-light interferometry system

## 3. RESULTS AND DISCUSSION

**3.1 Laser micromachining with laser beam scanning optical system**

**3.1.1: 103mm focal length objective: effects of pulse repetition rate**

The laser beam scanner having a 103 mm focusing objective was used to study the effects of varying the period between incident laser pulses on the micromachined cut. In this case the theoretical spot size is 13 μm and the depth of focus is approximately 50 μm. The processing parameters are listed in Table 1. The fluence was held constant at 42.2 J cm$^{-2}$ for each of the three cuts. The chosen pulse repetition rates were 5 kHz, 20 kHz, and 40 kHz. The laser beam scanning speeds were chosen to maintain a constant 0.71 laser beam pulse overlap for each of the three machining conditions. The period between incident laser pulses for the three different pulse repetition rates were 200, 50, and 25 microseconds respectively.

| Sample | Power @ Work Piece (W) | Pulse Repetition (kHz) | Scanning Speed (mm/sec) | Pulse Duration (ns) | Pulse Energy (uJ) | Energy Density (J/cm^2) | Avg Power Density (MW/cm^2) | Peak Power Density (GW/cm^2) |
|---|---|---|---|---|---|---|---|---|
| 1 | 0.28 | 5 | 18.75 | 30 | 56 | 42 | 0.2 | 1.4 |
| 2 | 1.12 | 20 | 75 | 35 | 56 | 42 | 0.8 | 1.2 |
| 3 | 2.25 | 40 | 150 | 45 | 56 | 42 | 1.7 | 0.9 |

Table 1. Processing parameters for cutting experiments using the 13 μm focal spot size (objective fl = 103 mm).

In the present case the buffer layer is composed of a 1-micron thick film of yttrium stabilized zirconia (YSZ) followed by a 20-nanometer thick cerium oxide ($CeO_2$) film. The SEM cross-sections presented in Figure 3 reveal the occurrence of a HAZ is most prominent with the 40 kHz condition and least prominent with the 5 kHz condition. The HAZ is most obviously characterized by the presence of deformation in the silver layer local to the groove as a result of melting and resolidification. Additionally, this effect is also observed in the nickel substrate. The region of melting and resolidification extends deeper into the substrate as the pulse repetition rate and pulse duration increase.

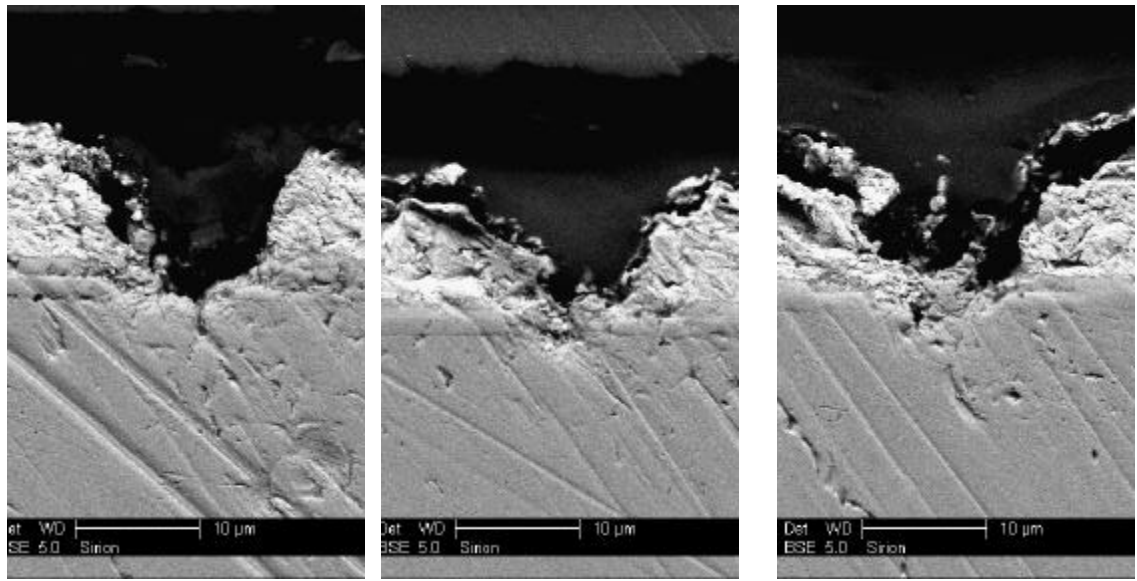

(a) 5 kHz (30 ns)   (b) 20 kHz (35 ns)   (c) 40 kHz (45 ns)

Figure 3. Groove cuts with constant pulse overlap and fluence. HAZ increases with decreasing period between laser pulses and increasing pulse duration. Pulse repetition rate (pulse duration) are given for each cut (a)-(c).

The occurrence of a HAZ within the metallic components of this heterostructure is likely attributed to greater thermal load on the work piece as a result of the longer laser pulse duration as well as a reduced cooling cycle associated with the increased pulse repetition rate. The lower 5 kHz pulse repetition rate has a pulse duration of 30 ns in comparison to 35 ns and 45 ns pulse duration at 20 kHz and 40 kHz respectively. Although no quantitative analysis has been performed, the reduced cooling period between laser pulses at higher pulse repetition rates likely contributes to an elevated preheated condition prior to the arrival of successive laser pulses.

The thermal diffusion length of the absorbing material is expressed in terms of the materials thermal diffusivity, $K_d$, and the incident laser pulse, duration, $t$:

$$d = 2\sqrt{K_d t} \qquad (1)$$

Increasing the laser pulse duration from 30 ns to 45 ns results in a 22.5% increase in the thermal diffusion length of each of the materials. This suggests that the HAZ should penetrate deeper into the surrounding material, which is consistent with the immediate experimental observation. In addition to increased pulse repetition rate and pulse duration, plasma heating can affect the presence of the HAZ observed in the nickel substrate and the silver coating. Plasma heating is prolonged because the thermal pumping time (pulse duration) is increased by 50 %. Exposure to the laser as a heat source allows more time for the energy to couple into the various thermal modes of the absorbing material. No observable thermal damage to the ceramic buffer and YBCO is observed in the optical and SEM images. The ceramic materials (YBCO and buffer) tend to vaporize without leaving evidence of melting and resolidification.

### 3.1.2 53mm focal length objective: effects of fluence

The laser beam scanner having a 53 mm focusing objective was used to study the effects of incident fluence on the micromachined groove. In this case the theoretical spot size is 6.5 µm and the depth of focus is approximately 25 µm. Although reducing the focal length of the focusing objective lens from 103 mm to 53 mm yields a smaller focal spot size, a reduced scan field and reduced depth of focus is incurred. A reduced depth of focus is beneficial in that control over the machining depth is increased; however, positioning resolution of the work piece becomes more critical. The benefits of reducing the focal spot size are primarily smaller machined feature size and a reduction of the necessary pulse energy. A two-fold reduction in the spot diameter results in a four-fold reduction in the pulse energy required to maintain a constant fluence. Thus, the average power may be reduced while maintaining a fluence that exceeds the damage threshold of the material for a given pulse repetition rate. Minimization of the average power generally leads to reduced HAZ provided the incident fluence exceeds the ablation threshold of the absorbing material.

| Sample | Power @ Work Piece (W) | Pulse Repetition (kHz) | Scanning Speed (mm/sec) | Pulse Duration (ns) | Pulse Energy (uJ) | Energy Density (J/cm^2) | Avg Power Density (MW/cm^2) | Peak Power Density (GW/cm^2) |
|---|---|---|---|---|---|---|---|---|
| 4 | 0.28 | 5 | 18.75 | 30 | 56 | 169 | 0.8 | 5.6 |
| 5 | 0.42 | 20 | 75 | 35 | 21 | 64 | 1.3 | 1.8 |
| 6 | 0.58 | 40 | 150 | 45 | 15 | 44 | 1.7 | 1.0 |

Table 2. Processing parameters for cutting experiments using the 6.5 µm focal spot size (fl = 53 mm)

The incident laser fluence was varied from 43.7 J cm$^{-2}$ to 169.4 J cm$^{-2}$ by decreasing the laser pulse energy for each of the three cuts. The processing parameters are listed in Table 2. The pulse repetition rate was increased in order for the laser to run in a stable mode of operation and maintain a reasonable average power output for low pulse energy. The pulse overlap was held constant at 0.41 for each of the three pulse repetition rates by adjusting the laser beam scanning speed.

In the present case the buffer layer is composed of a 1-micron thick film of yttrium stabilized zirconia (YSZ) followed by a 20-nanometer thick cerium oxide ($CeO_2$) film. The SEM cross-sections shown in Figure 4, indicate that increasing fluence has little adverse effect on the presence of HAZ and cut geometry. The cut depth increases slightly with decreased fluence and increased pulse repetition and pulse duration. Although comparable amounts of material are removed, the Sample 4 cut (169 J cm$^{-2}$, 5 kHz, 30 ns) appears to extend through more than half of the YBCO layer but not into the buffer layer. Whereas the Sample 5 cut (64 J cm$^{-2}$, 20 kHz, 35 ns) extends through the YBCO layer to the surface of the buffer layer and the Sample 6 cut (44 J cm$^{-2}$, 40 kHz, 45 ns) extends just into the buffer layer.

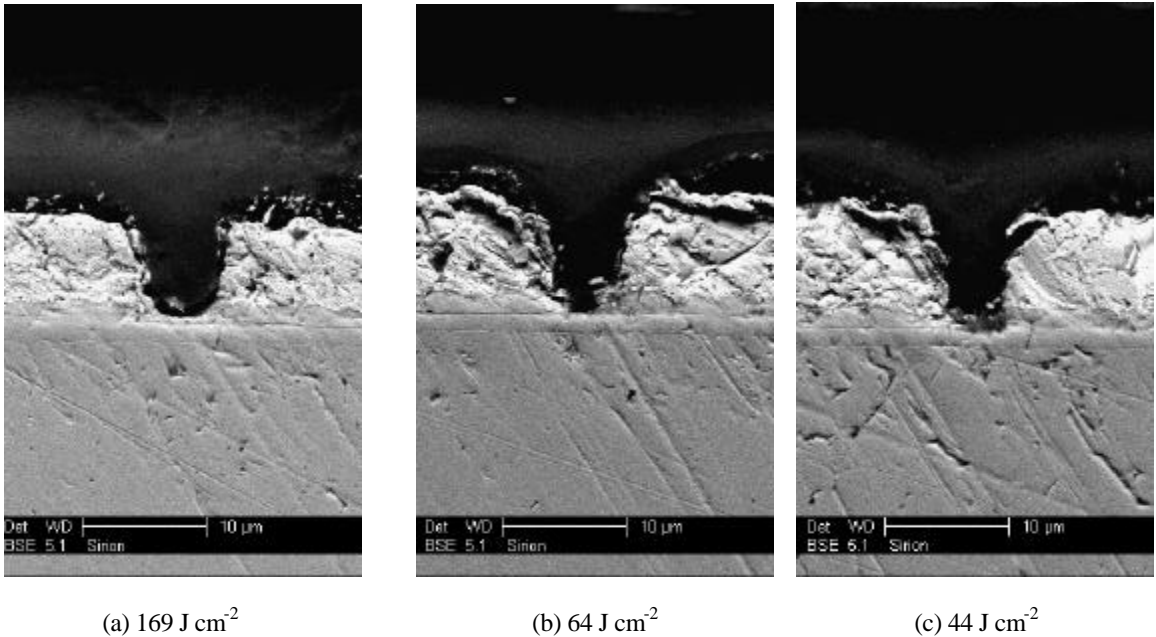

(a) 169 J cm$^{-2}$        (b) 64 J cm$^{-2}$        (c) 44 J cm$^{-2}$

Figure 4. Grooves cut at same pulse overlap and different incident fluence.

One might expect that a surplus of laser energy would promote the existence of a substantial HAZ. However, Samples 4-6 suggests that the combination of increased pulse repetition rate and pulse duration have a greater effect than increased fluence. As was observed in Samples 1-3, deformation of the silver layer along the cut walls becomes more pronounced as the pulse repetition rate and pulse duration increases even though the fluence is decreasing. Although metallic deformation is not as apparent when using the shorter focal length objective, it is difficult to attribute this directly to a smaller incident spot size because the pulse overlap was also reduced from 0.71 to 0.42 when using the 53 mm lens.

### 3.1.3 103mm and 53 mm focal length objective: direct comparison

For Samples 1 and 4 all processing parameters are constant including the incident pulse energy, pulse repetition rate, pulse duration, laser beam scanning speed, and as a result the amount of energy delivered to the work piece per unit length is constant. However, spatially the energy is more concentrated over a smaller area at the point of incidence when using the 53 mm lens. Additionally, the smaller spot size yields a reduced pulse overlap. SEM cross-sections are shown in Figure 5. The cut width is reduced proportionally with the spot size by factor of 2x from 10 µm to 5 µm. When using the 53 mm lens less damage to the silver coating is observed and the cut does not extend through the buffer layer into the nickel substrate.

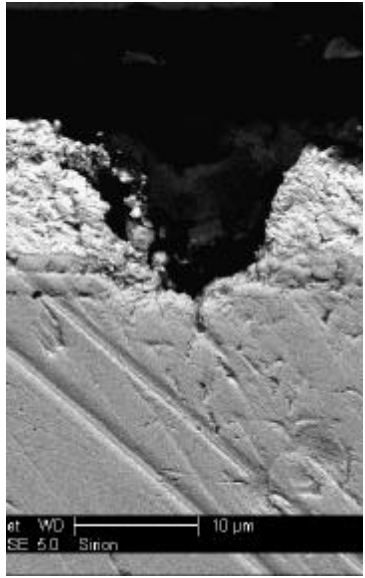 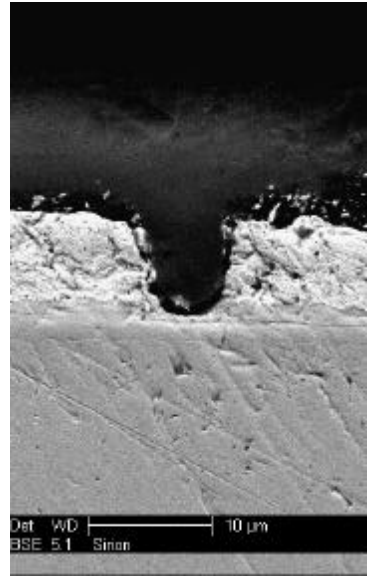

    (a) fl = 103 mm (13 μm spot size)      b) fl = 53 mm (6.5 μm spot size)

Figure 5. Under identical processing conditions, (a) Sample 1 cut and (b) Sample 4 cut.

### 3.2 Laser micromachining with fixed laser beam using a plano-convex objective

#### 3.2.1 The effects of laser beam scan speed: Buffer layer as an ablation barrier

The effects of varying the laser pulse overlap were investigated using the fixed 100 mm simple objective arrangement. A different coated conductor architecture was used for this experiment. In the present case the coated conductor includes a triple buffer layer comprised of $CeO_2$, YSZ, and $Y_2O_3$. Each buffer material layer is 75 nm thick and is deposited on a rolled and biaxially textured substrate (RABiTS) [3]. While holding all other laser processing condition constant, five different work piece translation speeds were used to effectively scan the stationary laser beam across the conductor surface. The translation speeds were 4.2, 8.3, 12.5, 14.2, and 16.7 mm $sec^{-1}$. The average incident power was 250 mW, the pulse repetition rate was 2.5 kHz, and the pulse duration was 30 nanoseconds. The processing parameters are summarized in Table 3.

| Sample | Power @ Work Piece (W) | Pulse Repetition (kHz) | Translation Speed (mm/sec) | Pulse Duration (uJ) | Pulse Energy (uJ) | Energy Density (J/cm^2) | Pulse Overlap | Cut Depth (um) |
|---|---|---|---|---|---|---|---|---|
| 7 | 0.25 | 2.5 | 4.2 | 30 | 100 | 44 | 0.902 | 11.9 |
| 8 | 0.25 | 2.5 | 8.3 | 30 | 100 | 44 | 0.804 | 7.9 |
| 9 | 0.25 | 2.5 | 12.5 | 30 | 100 | 44 | 0.706 | 5.15 |
| 10 | 0.25 | 2.5 | 14.2 | 30 | 100 | 44 | 0.667 | 4.1 |
| 11 | 0.25 | 2.5 | 16.7 | 30 | 100 | 44 | 0.608 | 4.4 |

Table 3. Processing parameters and cut depths for samples machined using simple objective (fl = 100 mm). All processing parameters held constant except work piece translation speed.

White Light interferometry was performed on each of the five cuts. Using this technique, cross sections of the cuts were determined and three-dimensional renderings were generated. A representative rendering is shown in Figure 6. Optical images of the same cut are shown in Figure 7. The cut depths were determined from the cross-section data and plotted as a function of pulse overlap as shown in Figure 8. The data show that as the translation speed is decreased from 16.7 mm sec$^{-1}$ to 14.2 mm sec$^{-1}$ the cut depth is essentially constant at 4 μm while further reducing the speed to 12.5 mm sec$^{-1}$ only increases the depth to 5 μm. However, further decreasing the scan speed to 8.3 mm sec$^{-1}$ and 4.2 mm sec$^{-1}$ shows a substantial increase in the cut depth to 8 μm and 12 μm respectively.

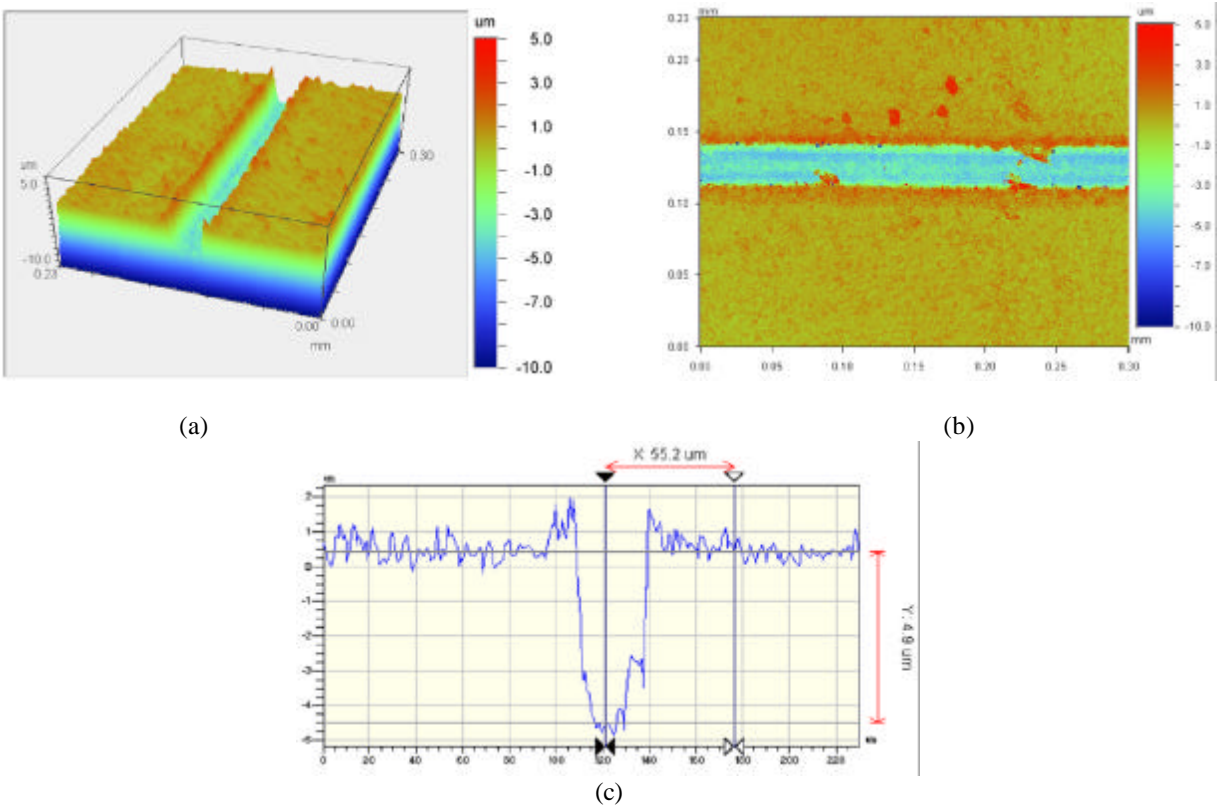

Figure 6. White light interferometric data for Sample 10 (14.2 mm sec$^{-1}$) including (a) 3-D rendering, (b) 2-D rendering, and (c) cross-section (not proportional).

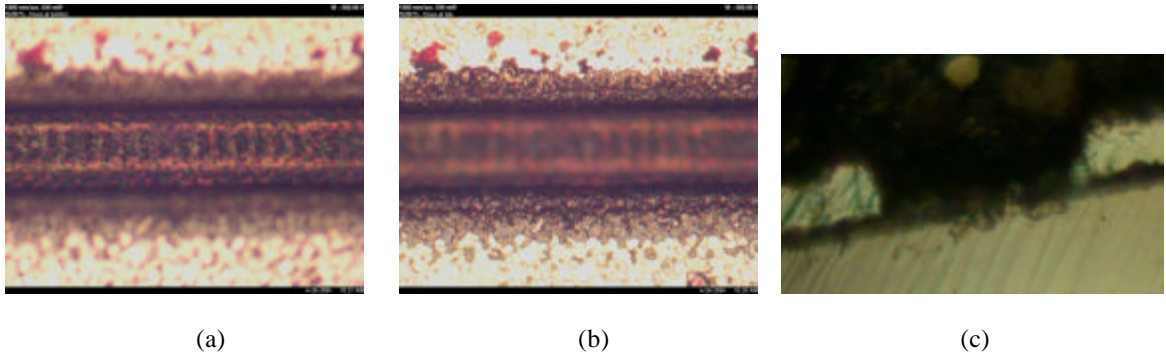

Figure 7. (a) 500x optical image focused at bottom of groove, (b) 500x optical image focused at top of groove, and (c) optical cross-section. Data shown is of Sample 10 (14.2 mm sec$^{-1}$).

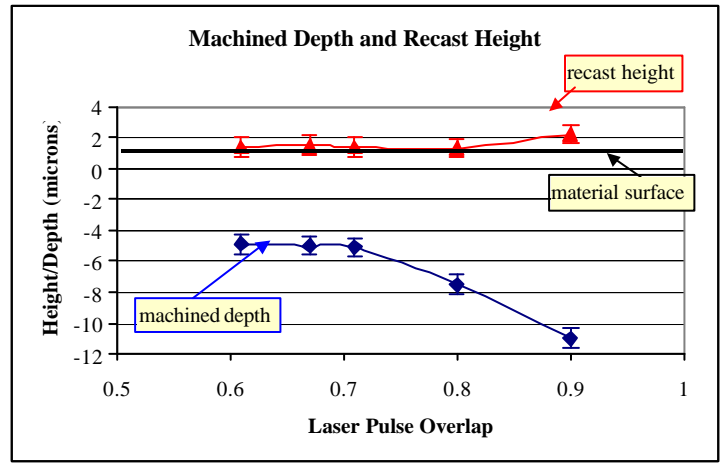

Figure 8. Cut depth and recast height as a function of laser beam pulse overlap (overlap decreases as translation speed increases) for Samples 7-11.

Considering that the interface depth between the silver coating and the buffer layer is approximately 5μm, it is likely the buffer layer is acting as an ablation barrier. The silver coating appears to easily ablate at the faster translation speeds. However, only after a significant decrease in the translation speed occurs does the cut penetrate the buffer layer. As the speed is further decreased the cut depth extends into the nickel substrate. Once the buffer layer is breached the cut depth increases significantly as the scan speed is further reduced. Optical cross-sections in Figure 9 reveal the evolution of a substantial nickel recast zone once the cut penetrates the nickel substrate, which extends to the top of the silver coating. The nickel recast provides a metallic coating effectively capping the edge of the YBCO and buffer layers. Additionally, an electrically conductive path may potentially exist between the silver coating, the nickel substrate, and the YBCO layer.

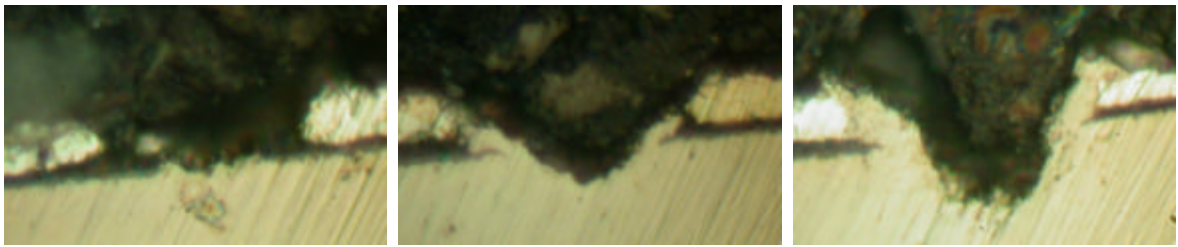

(a) 12.5 mm sec$^{-1}$      (b) 8.3 mm sec$^{-1}$      (c) 4.2mm sec$^{-1}$

Figure 9. Optical cross-sections of (a) Sample 9, (b) Sample 8, and (c) Sample 7.

### 3.3 Manufacturing Processing Rates

Ultimately material-processing rates will become a significant factor in the manufacturing scheme for coated conductor laser micromachining technology. The laser processing configurations considered in this study can potentially be integrated into an existing coated conductor manufacturing process. Coated conductor laser micromachining processing rates were calculated for both the fixed and scanning laser beam optical arrangements considered in this study. The calculated rates are summarized in Table 4. The rates are also given as a function of cut pitch along the length of a 1-cm wide coated conductor. The fastest processing rate occurs for the 53mm focal length laser beam scanner and 40 kHz pulse repetition rate. This configuration yields 5.4 meters of coated conductor in 1 hour for a cut pitch of 100 microns. The fixed laser beam configuration creating the nickel-recast barrier as described in Section 3.2.1 (Sample 7) has the lowest processing rate at 0.15 m hr$^{-1}$. The fixed laser beam configuration yielding the fastest processing rates were achieved when the cut extends just to the buffer layer (Sample 11). For this configuration the processing rate is 0.60 m hr$^{-1}$.

| Laser Micromachining Technique | Focal Length (mm) | Cut Type / Quality | Pulse Repetition Rate (kHz) | Scan Speed (mm/sec) | Cut Width (microns) | Cut Pitch (microns) | Coated Conductor Processing Rate (meters/hour) |
|---|---|---|---|---|---|---|---|
| Fixed Laser Beam | 100 | Thru buffer - Ni recast | 2.5 | 4.2 | 30 | 100 | 0.15 |
| | | | 2.5 | 4.2 | 30 | 50 | 0.08 |
| Fixed Laser Beam | 100 | To buffer - moderate | 2.5 | 16.7 | 30 | 100 | 0.60 |
| | | | 2.5 | 16.7 | 30 | 50 | 0.30 |
| Scanning Laser Beam | 103 | To buffer - moderate | 5 | 18.75 | 10 | 100 | 0.68 |
| | | | 5 | 18.75 | 10 | 50 | 0.34 |
| | | | 5 | 18.75 | 10 | 20 | 0.14 |
| Scanning Laser Beam | 53 | To buffer - best | 40 | 150 | 5 | 100 | 5.40 |
| | | | 40 | 150 | 5 | 50 | 2.70 |
| | | | 40 | 150 | 5 | 10 | 0.54 |

Table 4. Calculated laser micromachining rates for various optical conditions and cut densities.

## 4. SUMMARY

Both scanning and fixed optical systems were used to demonstrate the ability to laser micromachine grooves into the coated YBCO conductor - effectively striating the YBCO layer. Interesting effects were observed as a result of the unique combination of the metallic substrate, silver coating, and the intermediate ceramic thin films. A heat-affected zone was shown to exist within the nickel substrate and silver coating under specific laser processing conditions. In the case of the laser beam scanning optical configuration, the conditions contributing most to a HAZ were the longer pulse duration and shorter cooling cycles associated with the higher pulse repetition rates (upwards to 40 kHz). Although also observed to a minimal degree using the shorter 53 mm focal length objective, the presence of the HAZ was most pronounced when using the longer 103 mm focal length objective. Additionally, when using the shorter 53 mm focal length objective, no advantage was observed when increasing the laser pulse energy above 15 µJ (44 J cm$^{-2}$).

Additionally, the experiments performed using the fixed optical system showed the buffer layer as being particularly resistant to laser ablation. Using the simple plano-covex focusing lens at an incident fluence of 44 J cm$^{-2}$, the cut depth was constant at the buffer layer as the translation speed was decreased. Only when the translation speed was substantially decreased did the laser beam penetrate the buffer layer after which a strong dependence on cut depth on translation speed was observed. Further, as the cut is machined deeper into the nickel substrate, a recast zone is observed which extends up the side of the cut reaching above the silver coating - effectively capping the YBCO layer and also potentially creating an electrically conductive path.

Finally, processing rates were calculated as a function of cut density for both the fixed and scanning laser beam optical arrangements considered in this study. The calculations show that the shorter 53 mm focal length laser beam scanner and 40 kHz pulse repetition rate configuration achieve the highest processing rates: 5.4 m hr$^{-1}$ for 100 micron cut pitch along the 1-cm wide coated conductor. These same conditions produce the smallest cut width and minimal HAZ. The results of this study suggest that the optimal processing arrangement should utilize a high pulse repetition rate (>40 kHz), short laser pulse duration (<30 nsec), low pulse energy (~15 µJ), rapid laser beam scanning speeds (>150 mm sec$^{-1}$), and a short focal length objective (~50 mm) achieving the smallest focal spot size (~6 µm).

## ACKNOWLEDGMENTS

The authors would like to thank the Superconductivity Group, especially P. N. Barnes and C.E. Oberly, of the Propulsion Directorate, Air Force Research Laboratory, and L.R. Dosser of the Mound Laser and Photonics Center for financial and scientific support of this work. Work by G.A.L. was performed while he held a National Research Council Senior Research Associateship Award at the Air Force Research Laboratory.